\documentclass[12pt,preprint]{aastex}

\begin{document}

\title{Accretion-Driven Turbulence and the Transition to Global
Instability in Young Galaxy Disks}

\author{Bruce G. Elmegreen}
\affil{IBM Research Division, T.J. Watson Research Center, P.O. Box
218, Yorktown Heights, NY 10598} \email{bge@watson.ibm.com}

\author{Andreas Burkert}
\affil{University Observatory Munich (USM), Scheinerstrasse 1, 81679
Munich, Germany, Max-Planck-Fellow} \email{burkert@usm.uni-muenchen.de}

\begin{abstract}
A simple model of gas accretion in young galaxy disks suggests that
fast turbulent motions can be driven by accretion energy for a time
$t_{\rm acc}\sim2\left(\epsilon^{0.5} GM^2/\xi V^3\right)^{0.5}$ where
$\epsilon$ is the fraction of the accretion energy going into disk
turbulence, $M$ and $V$ are the galaxy mass and rotation speed, and
$\xi$ is the accretion rate.  After $t_{\rm acc}$, accretion is
replaced by disk instabilities as a source of turbulence driving, and
shortly after that, energetic feedback by young stars should become
important. The star formation rate equilibrates at the accretion rate
after 1 to 2 $t_{\rm acc}$, depending on the star formation efficiency
per dynamical time. The fast turbulence that is observed in high
redshift starburst disks is not likely to be driven by accretion
because the initial $t_{\rm acc}$ phase is over by the time the
starburst is present. However, the high turbulent speeds that must have
been present earlier, when the observed massive clumps first formed,
could have been driven by accretion energy.  The combined observations
of a high relative velocity dispersion in the gas of $z\sim2$ clumpy
galaxies and a gas mass comparable to the stellar mass suggests that
either the star formation efficiency is fairly high, perhaps $10\times$
higher than in local galaxies, or the observed turbulence is powered by
young stars.
\end{abstract}


\section{Introduction}\label{intro}

Deep surveys have detected Milky-Way-size galaxy disks at redshift $z
\sim 2$ that contribute a large fraction of the stellar mass density
and have star formation rates of order 100 $M_{\odot}$ yr$^{-1}$ (Daddi
et al. 2004, 2007; F\"orster-Schreiber et al. 2006; Erb et al. 2006a,b;
van Dokkum et al. 2006). They appear to be highly turbulent with gas
velocity dispersions of 40-80 km s$^{-1}$ (F\"orster-Schreiber et al.
2006, 2009; Genzel et al. 2006, 2008; Cresci et al. 2009). They form
stars in a small number of clumps with masses of around the turbulent
Jeans mass, which is $\sim10^8\;M_\odot$ (Elmegreen \& Elmegreen 2005;
Bournaud et al. 2007, 2008). They are also gas-rich, with gas-to-star
mass fractions of order 30-60\% (Tacconi et al. 2009). These features
are in contrast to present-day disk galaxies that are relatively
quiescent by comparison, having turbulent velocities of order 10 km
s$^{-1}$ (Dib et al. 2006), star-forming complexes typically smaller
than $10^6\;M_\odot$, and star formation rates of order 1-3 $M_{\odot}$
yr$^{-1}$.

The presence of gas does not necessarily imply star formation. Certain
conditions have to be met before gas can become unstable to form
self-gravitating clouds (e.g., Toomre 1964; Goldreich \& Lynden-Bell
1965), and other conditions are required before these clouds can cool
enough to condense into stellar objects (e.g., Elmegreen \& Parravano
1994; Schaye 2004; Krumholz, McKee, \& Tumlinson 2009; Prochaska \&
Wolfe 2009). These conditions have the effect of delaying the onset of
star formation in a young galaxy, allowing the gas mass to build up
without attrition until it reaches a tipping point, after which the
conditions change and star formation is fast. The observed
star-formation rate of $100\;M_\odot$ yr$^{-1}$ corresponds to nearly a
whole galaxy disk of gas, several $\times 10^{10} M_{\odot}$, being
converted into stars in only a few orbit times, which is several
$\times 10^{8}$ yrs. If the gas accretion rate is comparable to this
star formation rate, then the initial build up of a disk is fast and
star formation has to be very efficient, with a shorter gas consumption
time than in local galaxies by a factor of $\sim5$.  If the accretion
rate is less than the star formation rate, then there was probably a
prior phase when most of the gas assembled before the star formation
began.

The properties of galaxy disks during the initial accretion and
starburst phase are not understood.  The large masses of star-forming
clumps suggest that the gas was highly turbulent before the clumps
formed. In this case, the primary source of turbulent energy would seem
to be the accretion itself (e.g., Elmegreen \& Elmegreen 2005;
F\"orster-Schreiber et al. 2006; Genzel et al. 2008; Dekel et al. 2009;
Khochfar \& Silk 2009; Klessen \& Hennebelle 2009). After the disk
becomes unstable, further energy can come from disk self-gravity
(Burkert et al. 2009; Bournaud et al. 2009). Here we investigate the
duration of the initial turbulent phase when most of the energy comes
from cosmological accretion, and we consider the cooling that leads to
instabilities and star formation. We write equations for energy, mass,
and star formation in dimensionless form, and then normalize these
equations in a way that is independent of the gas accretion rate and
the efficiency of conversion from galaxy potential energy to disk
turbulent energy.  The equations are in Section 2, the solutions are in
Section 3, and a comparison with observations is in Section 4.

\section{An Accretion-Dissipation Equation}
\label{acc-diss}

We simplify a galaxy during the short phase of initial accretion by
considering a constant gas accretion rate $dM_{\rm disk}/dt \equiv \xi$
into a disk with fixed radius $R$ within a fixed potential $V^2$. The
energy accretion rate is approximately $\xi$ multiplied by the depth of
the potential well, $V^2$, assuming the thermal speed in the accretion
flow is small compared with $V$. We let $\epsilon$ be the efficiency of
conversion of accretion energy into turbulent energy, $E$, in the disk.
If $\epsilon<1$, then some of the potential energy from accretion is
lost to shocks and thermal radiation before it drives disk turbulence.
The rate of input of turbulent energy is $\epsilon \xi V^2$. If dark
matter accretes along with the gas, then $V^2$ will increase with time.
This is considered in Section \ref{time}.

An accreting disk dissipates turbulent energy at a rate approximately
equal to the current energy divided by the perpendicular crossing time
(e.g., Mac Low et al. 1998; Stone, Ostriker \& Gammie 1998). This
crossing time is $t_{\rm perp}=H/\sigma$ for disk scale height
$H=\sigma^2/\pi G\Sigma_{\rm disk} = \sigma^2R^2/GM_{\rm disk}$. The
turbulent speed is denoted by $\sigma=\left(2E/3M_{\rm
gas}\right)^{0.5}$ and the accreted disk mass is $M_{\rm disk}={\dot
M_{\rm disk}}t$. Thus the dissipation rate is $E\sigma/H=G (1.5M_{\rm
gas}E)^{0.5} M_{\rm disk} R^{-2}$. Putting accretion and dissipation
together, the time-dependent equation for turbulent energy is

\begin{equation}
{{dE}\over{dt}}=\epsilon \xi V^2 - \left(1.5M_{\rm gas}E \right)^{0.5}
\left( \frac{GM_{\rm disk}}{R^2}\right) -1.5S\sigma^2.
\label{edot}
\end{equation}

\noindent The last term is the energy lost by locking up interstellar
matter with velocity dispersion $\sigma$ into stars at a rate $S$. This
counts for energy lost from the gas phase; i.e., the total gas energy
decreases from this effect at a rate equal to the rate of conversion of
gas into stars multiplied by the energy content of that gas.

Star formation converts disk gas into stars. We take the disk gas mass
to be $M_{\rm gas}$ and the disk star mass to be $M_{\rm stars}$, so
that $M_{\rm gas}+M_{\rm stars}=M_{\rm disk}$. No gas or stars are
assumed to be ejected from the disk. The star formation rate is taken
to be $S_0$ times the gas mass $M_{\rm gas}$ multiplied by the growth
rate of a gravitational instability, which is $\omega=\pi G\Sigma_{\rm
gas}/\sigma$ in the linear instability theory. The coefficient $S_0$ is
an efficiency of star formation in a dynamical time; it is observed to
be $S_0\sim0.01$ on a variety of scales (Kennicutt 1998; Krumholtz \&
Tan 2007) reflecting the small fraction of cloud mass that is dense
enough to form stars and the rapid disruption of star-forming cores.
The total star formation rate in the galaxy is then
\begin{equation}
S=S_0 \frac{G M_{\rm gas}^2}{R^2\sigma}= S_0 \sqrt{1.5} \times \frac{G
M_{\rm gas}^{2.5}}{R^2E^{0.5}}. \label{S}
\end{equation}
Note that accretion followed by disk shrinkage at constant $M_{\rm
gas}$ causes the gas surface density to increase and the radius $R$ in
this equation to decrease.  Shrinkage is expected if gravitational
instabilities drive turbulence, because then turbulent energy is taken
from gravitational binding energy, and the dissipation of this energy
requires settling of the gas to a smaller radius. In the following
discussions, the length scales will be normalized to the galaxy radius
$R$. Then disk shrinkage should be viewed as a larger effective value
of $S_0$, according to equation (\ref{S}).

Because of star formation, the gas mass in the disk varies with time as
\begin{equation} {{dM_{\rm gas}}\over{dt}}=\xi-S, \label{mgas} \end{equation}
and the star mass varies as
\begin{equation}{{dM_{\rm stars}}\over{dt}}=S.\label{mstars}\end{equation}.

We now normalize the important physical parameters, writing time in
units of $R/V$, gas or disk mass in units of galaxy mass, $M_0$, and
energy in units of $M_0V^2$.  Geometric terms of order unity are
ignored, as are variations throughout the disk. There are two
parameters in this model, $\xi$ and $\epsilon$. We can absorb them into
our new variables to write the equations independent of them. Then we
get general solutions to the equations that are independent of all
parameters.  The gas accretion rate $\xi$ is dimensional, so we
introduce the dimensionless accretion rate, $A=\xi R/(VM_0)$, and
define:
\begin{eqnarray*}
M_0 & = & \sqrt{\frac{2}{3}}\frac{RV^2}{G}\;;\; {E^\prime} =
{{E}\over{A^{0.5}\epsilon^{1.25}M_0V^2}}\;;\;  {t^\prime}  =
{{tA^{0.5}V}\over{\epsilon^{0.25}R}}\;;\;\\
{M^\prime_{\rm disk}} & = & {{M_{\rm
disk}}\over{A^{0.5}\epsilon^{0.25}M_0}}\;;\;  {M^\prime_{\rm gas}} =
{{M_{\rm gas}}\over{A^{0.5}\epsilon^{0.25}M_0}}\;;\;
{M^\prime_{\rm stars}} = {{M_{\rm stars}}\over{A^{0.5}\epsilon^{0.25}M_0}}\;;\; \\
{\sigma^\prime} & = & {{\sigma}\over{\epsilon^{0.5}V}}  =
\left({{2E^\prime}\over{3M^\prime_{\rm gas}}}\right)^{0.5}\;;\;
{S^\prime} = {{S}\over{\xi}}={{SR}\over{AVM_0}}.
\end{eqnarray*}

\noindent Then equations (\ref{edot})-(\ref{mstars}) can be written
\begin{equation}
{{d{E^\prime}}\over{d{t^\prime}}}=1-M^\prime_{\rm
disk}\left({E^\prime}{M^\prime_{\rm
gas}}\right)^{0.5}-{S^\prime}{E^\prime/}{M^\prime_{\rm
gas}}\label{edotprime}\end{equation}
\begin{equation}
{S^\prime}=S_0 \left({M^\prime_{\rm
gas}}^{5}/{E^\prime}\right)^{0.5}\end{equation}
\begin{equation}
{{{dM^\prime}_{\rm disk}}
\over{d{t^\prime}}}=1\label{mdiskdotprime}\end{equation}
\begin{equation}{{{dM^\prime}_{\rm
gas}} \over{d{t^\prime}}}=1-{S^\prime}\label{mdotprime}\end{equation}
\begin{equation}{{d{M^\prime}_{\rm stars}}\over{d{t^\prime}}}= {S^\prime}
\label{mdotstarsprime}.
\end{equation}

We also define the useful quantities
\begin{eqnarray*}
Q^\prime & = &
QA^{0.5}/\epsilon^{0.25}=3^{0.5}\sigma^\prime/M^\prime_{\rm
gas}\\
M^\prime_{\rm Jeans} & = & M_{\rm Jeans}A^{0.5}/(\epsilon^{1.75}M_0)=
1.5\pi(\sigma^\prime)^4/M^\prime_{\rm gas}\\
\omega^\prime & = & \omega\epsilon^{0.25}R/(A^{0.5}V)=
(2/3)^{0.5}(M^\prime_{\rm gas}/\sigma^\prime)
\end{eqnarray*}
Here $Q=\kappa\sigma/(\pi G \Sigma)$ is the Toomre instability
parameter for epicyclic frequency $\kappa=2^{0.5}V/R$ in a flat
rotation curve and mass column density $\Sigma=M_{\rm gas}/(\pi R^2)$.
The unstable Jeans mass is $M_{\rm Jeans}=\sigma^4/(G\Sigma)$.  The
growth rate $\omega$ was given above. All primed quantities are
independent of $\xi$ and $\epsilon$. Note also that $\omega^\prime
t^\prime=\omega t$.

Equations (\ref{edotprime})-(\ref{mdotstarsprime}) were solved
numerically. The initial conditions are ${M^\prime}_{\rm gas}=0$,
${M^\prime}_{disk}=0$, and ${E^\prime}=0$. The solution is initially
dominated by the first terms in equations (\ref{edotprime}) and
(\ref{mdotprime}) so that ${E^\prime}\propto {M^\prime}_{\rm gas}$ at
first, giving $\sigma^\prime=1$ before dissipation becomes important.
In this regime, disk turbulence is fast and driven only by gas
accretion.

Physical quantities can be determined from the normalized variables if
we specify the corresponding galactic environment. Typical
high-redshift star-forming galaxies are characterized by $R \approx $
10 kpc and $V \approx $ 220 km s$^{-1}$, which give $M_0 \approx
9\times10^{10} M_{\odot}$ and $R/V=45$ Myr. If their accretion rates
are 50\% of the mass $M_0$ in one orbit time, $2\pi R/V=280$ Myr, then
$\xi=S/S^\prime\sim160\;M_\odot$ yr$^{-1}$ and $A=1/(4\pi)=0.08$. The
efficiency of conversion of infall energy into disk turbulent energy is
not well known. Klessen \& Hennebelle (2009) suggest $\epsilon$ is a
few percent to 10\%.  If we adopt $\epsilon =0.1$, then $t/{t^\prime} =
90$ Myrs, $M_{\rm gas}/{M^\prime_{\rm gas}} = 1.4 \times 10^{10}
M_{\odot}$, $E/{E^\prime} = 1.4 \times 10^{57}$ erg, and
$\sigma/{\sigma^\prime} = 70$ km s$^{-1}$.

\section{Results}

\begin{figure}\epsscale{1}
\plotone{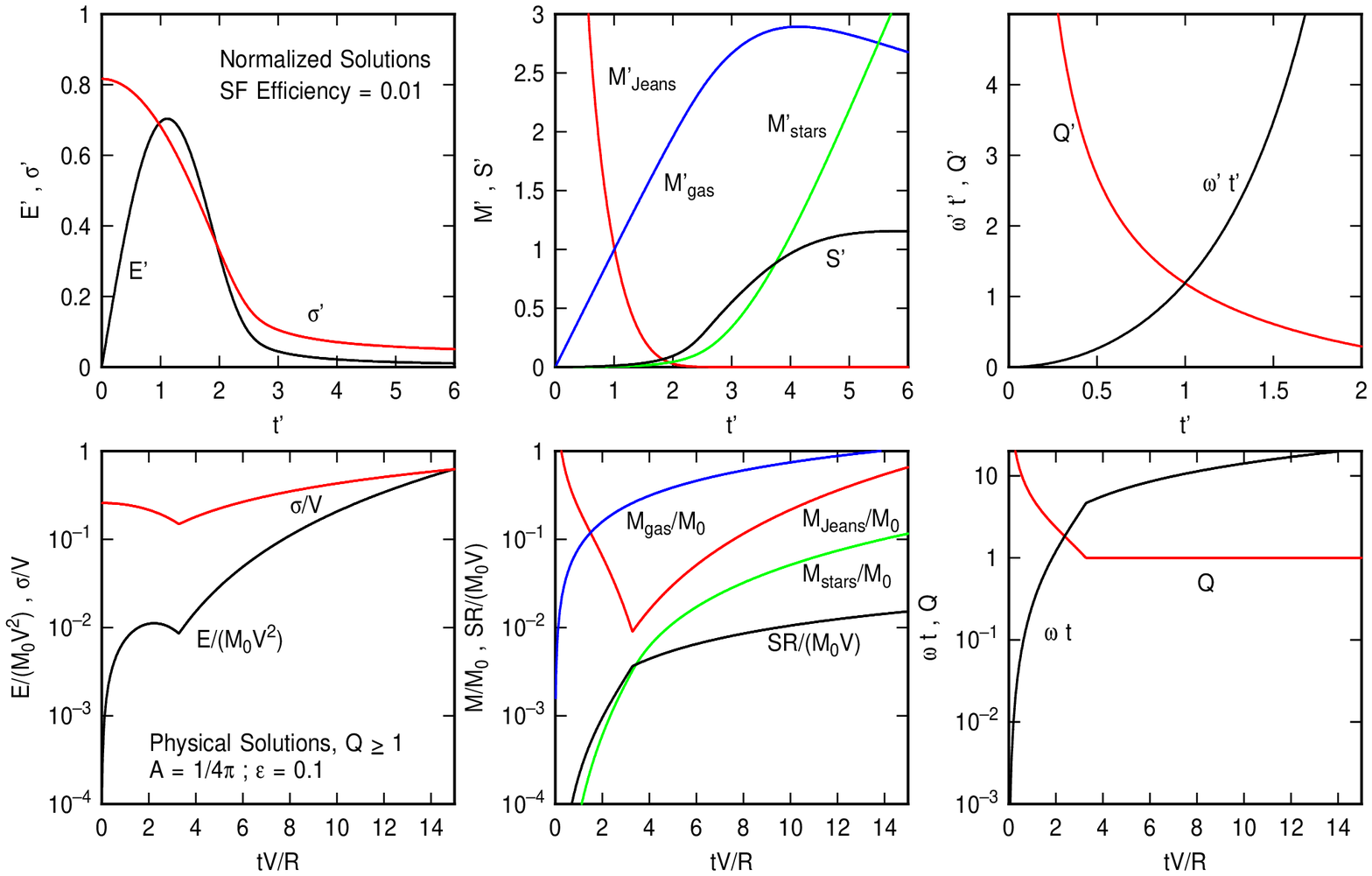} \caption{Normalized solutions to the equations of
accretion, dissipation, and star formation are on the top, and physical
solutions with $Q$ regulated to be larger than or equal to 1 are on the
bottom, both for the case where the star formation efficiency, $S_0$,
equals $0.01$. The left panels show energy and velocity dispersion, the
middle panels show masses and the star formation rate, the right panels
show the stability parameter and the product of the growth rate and
time. }\label{infall2f}\end{figure}

The top panels of Figure \ref{infall2f} show solutions to the
normalized equations as functions of normalized time for star formation
efficiency $S_0=0.01$. The turbulent energy in the disk increases
linearly with time at first and then decreases when the dissipation
rate becomes larger than the energy accretion rate at about
$t^\prime=1$. The normalized velocity dispersion in the disk starts at
a value of unity and decreases when dissipation becomes important. The
epoch of significant accretion-driven turbulence is $t^\prime<2$.

The unstable mass in a turbulent disk is the Jeans mass, $M_{\rm
Jeans}\sim\sigma^4/G^2 \Sigma$, which in dimensionless units is
$M^\prime_{\rm Jeans}=1.5\pi\left(\sigma^\prime\right)^4/M^\prime_{\rm
gas}$ written above. The ratio of the Jeans mass to the gas mass is
\begin{equation}
M_{\rm jeans}/M_{\rm gas}=(M^\prime_{\rm Jeans}/M^\prime_{\rm gas})
\epsilon^{1.5}/A.\end{equation} Initially $M_{\rm Jeans}>>M_{\rm gas}$
and no gravitational instabilities are possible. When $M_{\rm Jeans}
\lesssim0.5M_{\rm gas}$, instabilities can begin; i.e., each half of
the disk can clump into a separate cloud. This condition corresponds to
$M^\prime_{\rm Jeans}/M^\prime_{\rm gas}\lesssim 0.5A/\epsilon^{1.5}$.
For $A=0.08$ and $\epsilon=0.1$ as above, this becomes $M^\prime_{\rm
Jeans}/M^\prime_{\rm gas}\lesssim1.26$.  In Figure \ref{infall2f}, this
occurs at $t^\prime\gtrsim$0.94 or $t>85$ Myr with these $A$ and
$\epsilon$.

The time evolution of the normalized instability parameter $Q^\prime$
is shown on the top right in Figure \ref{infall2f}. It starts high and
decreases as $\sigma^\prime$ decreases and $M^\prime_{\rm gas}$
increases. If significant instabilities occur when $Q\lesssim1$, then
this corresponds to $Q^\prime\lesssim A^{0.5}/\epsilon^{0.25}$.  With
$A=0.08$ and $\epsilon=0.1$, this requires $Q^\prime<0.5$, and it
occurs in Figure \ref{infall2f} at $t^\prime>1.64$, or $t>148$ Myr.

\begin{figure}\epsscale{.7}
\plotone{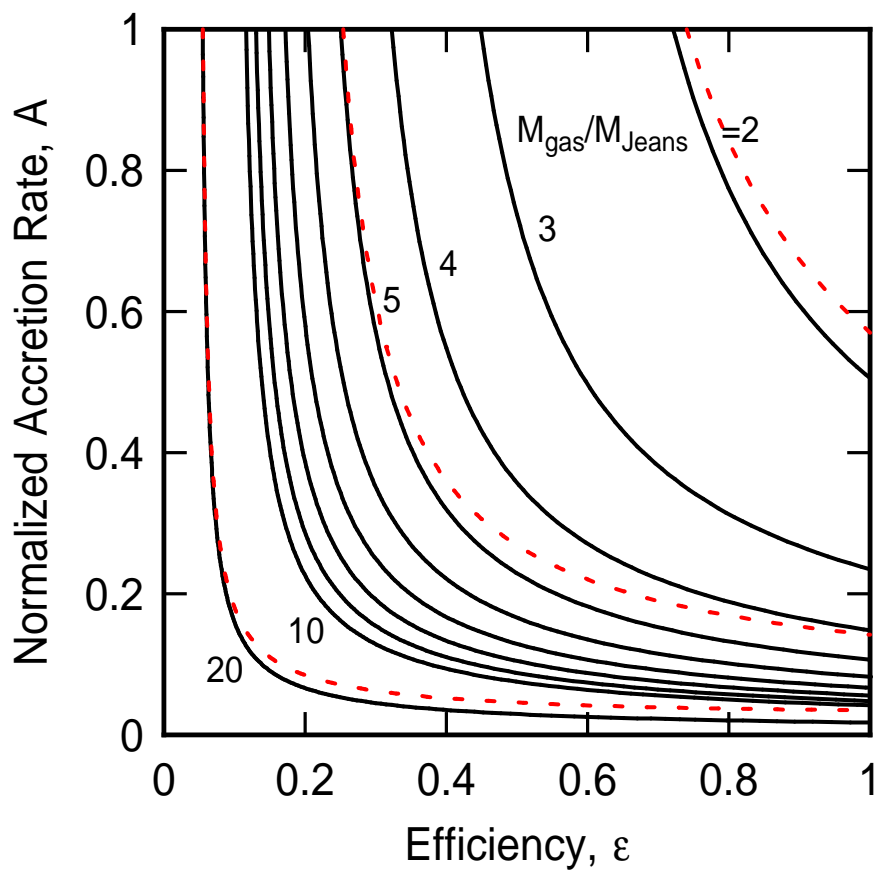} \caption{The curves show the ratio of the gas mass to
the turbulent Jeans mass as a function of the normalized accretion
rate, $A=\xi R/(VM_0)$ for physical rate $\xi$, and the efficiency
$\epsilon$ for the conversion of accretion energy into turbulence. The
solid black curves assume a star formation efficiency of $S_0=0.01$ and
the three red dashed curves assume $S_0=0.1$ for the $M_{\rm
gas}/M_{\rm Jeans}=2$, 5, and 20 cases.  This ratio of masses should be
about equal to the number of giant clumps formed in the first epoch of
instabilities. }\label{infall6}\end{figure}

The number of giant clouds that form from the instability equals about
$M_{\rm gas}/M_{\rm Jeans}$. Note that $M_{\rm gas}/M_{\rm Jeans}$
exceeds the minimum likely value, $\sim2$, when the disk is still
stable, i.e. when $Q>1$, for these $A$ and $\epsilon$. Thus as soon as
the disk becomes unstable, $Q<1$, there is enough mass in it to form
clouds. The condition for instability is therefore the Toomre $Q<1$
condition, rather than the Jeans minimum-mass condition, $M_{\rm
gas}/M_{\rm Jeans}>2$.  For most $A$ and $\epsilon$, this is the case.
Figure \ref{infall6} shows with the top right curve the values of $A$
versus $\epsilon$ for which $M_{\rm gas}/M_{\rm Jeans}=2$ and $Q=1$
occur at the same time. A star formation efficiency $S_0=0.01$ is
assumed. Values of $A$ and $\epsilon$ below and to the left of this
curve have $M_{\rm gas}/M_{\rm Jeans}>2$ when the disk is still stable.
Thus all of this lower-left region in $A-\epsilon$ space begins star
formation because of the Toomre condition, rather than the Jeans
condition. Above and to the right of the $M_{\rm gas}/M_{\rm Jeans}=2$
curve, the Jeans condition determines the onset of star formation. The
other black curves in Figure \ref{infall6} correspond to values of
$M_{\rm gas}/M_{\rm Jeans}$ when the first point of instability occurs,
at $Q=1$, also for $S_0=0.01$.  Loosely interpreted, these curves give
the number of giant clouds that form in the disk. The red dashed curves
are for a higher efficiency, $S_0=0.1$, which does not change the
values much.

\begin{figure}\epsscale{.7}
\plotone{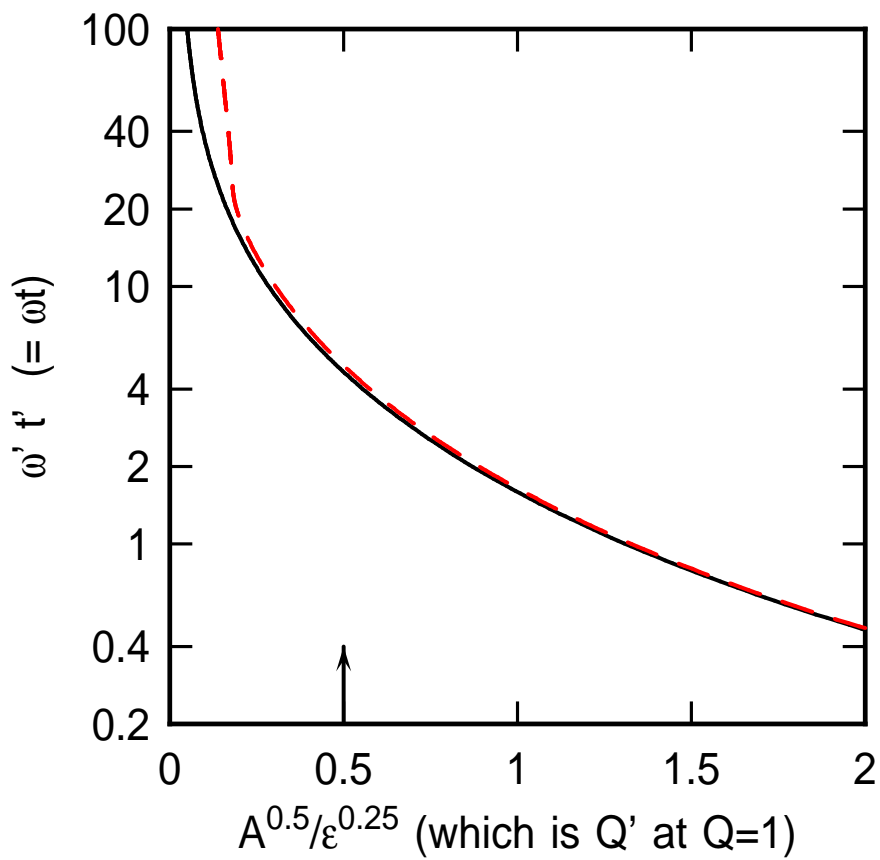} \caption{The product of the normalized growth rate and
the normalized time, which is the same as the product of the physical
growth rate and time, versus a combination of $A$ (accretion rate) and
$\epsilon$ (turbulence driving efficiency) that appears in the
normalized instability parameter $Q^\prime$.  For $A=0.08$ and
$\epsilon=0.1$, as discussed in the text, this combination equals 0.5,
which is shown by the arrow. The large values of $\omega^\prime
t^\prime$ suggest that the instability should be rapid once the disk
goes unstable. The solid black curve is for $S_0=0.01$ and the dashed
red curve is for $S_0=0.1$.}\label{infall7}\end{figure}

The top right panel of Figure \ref{infall2f} shows the product of the
time and the instability growth rate. The product $\omega^\prime
t^\prime$ increases and exceeds 1 when $t^\prime>0.93$. It is generally
large by the time $Q<1$.  Note that $\omega^\prime t^\prime=\omega t$,
independent of $A$ and $\epsilon$, so that when $\omega^\prime
t^\prime>1$, the unstable growth occurs quickly compared to the age of
the galaxy.  Such quick growth is generally applicable in our models.
The value of $\omega^\prime t^\prime$ at the onset of instability
($Q=1$) is shown in Figure \ref{infall7}. The solid black curve uses
$S_0=0.01$ and the dashed red curve uses $S_0=0.1$. Both assume $Q$
decreases continuously with turbulent dissipation, i.e., without a
$Q=1$ floor. What is plotted is $\omega^\prime t^\prime$ versus
$A^{0.5}/\epsilon^{0.25}$, which is the value of $Q^\prime$ at $Q=1$;
this is a normalized curve, independent of $A$ and $\epsilon$. For
$A^{0.5}/\epsilon^{0.25}<1.31$ in the $S_0=0.01$ case, $\omega^\prime
t^\prime>1$ at the onset of instabilities. For fiducial $A=0.08$ and
$\epsilon=0.1$, $A^{0.5}/\epsilon^{0.25}=0.5$ and $\omega^\prime
t^\prime=4.65$, as indicated in the figure by the arrow. The rapid
turnup in the $S_0=0.1$ case is because $Q^\prime$ levels off faster at
late times than in the $S_0=0.01$ case as a result of the decrease in
$M_{\rm gas}$ that follows from the higher star formation rate with
$S_0=0.1$.

The top center panel in Figure \ref{infall2f} shows that the normalized
star formation rate approximately equals the gas accretion rate
($\equiv1$ for our normalized quantities) after the time
$t^\prime\sim3$, at which point the ratio of the turbulent dispersion
speed to the orbit speed has decreased to $\sim0.1$ in this $S_0=0.01$
case. Star formation increases so rapidly after $t^\prime\sim2$ that
the stellar disk mass soon exceeds the gas mass, which happens at
$t^\prime=5.5$ in Figure \ref{infall2f}. Then the disk enters a
near-steady state with a powerful starburst lasting as long as the high
accretion continues. Disk self-gravity should contribute to the
turbulent energy after $t^\prime\sim1$, when disk instabilities become
active, and star formation should contribute to the turbulent energy
after $t^\prime\sim2$, when the starburst begins.

A problem with these solutions is that $\sigma$ becomes small before
the starburst begins and before $M_{\rm gas}\sim M_{\rm stars}$.
Observations suggest the opposite, that $\sigma$ is a large fraction of
the rotation speed in bursting galaxies with $M_{\rm gas}\sim M_{\rm
stars}$. This problem occurs for pure accretion-driven turbulence,
which always dies out after $t^\prime\sim1$. In a real galaxy,
gravitational instabilities and star formation drive turbulence and
prevent $Q$ from dropping much lower than 1 (Burkert et al. 2009;
Bournaud et al. 2009; Dekel, Sari, \& Cervino 2009). We simulate this
here by allowing $Q$ to decrease as $\sigma$ decreases and $M_{\rm
gas}$ increases, but when $Q=1$, we stop the decrease in $\sigma$ and
set $\sigma=\pi G \Sigma/\kappa$, which keeps $Q=1$. This means that
$\sigma$ decreases at first because of the dissipation of accretion
energy, but then it increases in direct proportion to the gas mass in
order to keep $Q=1$.

The bottom panels of Figure \ref{infall2f} show the same variables
again as in the top panel, but now with this $Q \geq1$ constraint, and
also plotted with explicit evaluation of the $A$ and $\epsilon$
dependencies, so that the variables are physical with normalization to
$M_0$, $V$, and R. The curve for $\sigma/V$ shows the effect just
discussed: the dispersion decreases at first and then increases once
$Q$ reaches 1 in order to keep $Q=1$, as indicated in the right panel
where $Q$ itself is plotted. The energy decreases at first, from
dissipation, but then increases from the assumed gravitational
instabilities. While this situation is closer to reality than an
ever-cooling disk, the high $\sigma$ that results keeps the star
formation rate low. Then the stellar mass builds up very slowly and is
always much less than the gas mass over the timespan plotted. Note that
$SR/(M_0V)$ in the figure barely gets above $0.01$, whereas the
accretion rate in these same units, $\xi R/(M_0V)=A$ is
$1/(4\pi)=0.08$. Thus the star formation rate has not yet equilibrated
to the accretion rate in this time.  Evidently, forcing $Q\geq1$ makes
$\sigma/V$ more reasonable for high redshift disks, but then star
formation is too slow and the stellar mass does not readily build up to
equal the gas mass.

To fix this second problem, we have to increase the efficiency of star
formation, $S_0$. Figure \ref{infall2g} shows the same normalized
variables in the top panel as in Figure \ref{infall2f}, again without
$Q$ restrictions,  and the same physical variables in the bottom panel
as in Figure \ref{infall2f}, again with $Q\geq1$, but this time $S_0$
is 10 times larger, $0.1$.  Now the star formation rate gets large
quickly, by $t^\prime\sim2$, it saturates to the gas accretion rate
even in the high-$\sigma$ case (bottom panels), and the stellar mass
becomes equal to the gas mass within the plotted physical time range,
at $tV/R=11.56$.

The time when $M_{\rm gas}=M_{\rm star}$ depends on the star formation
efficiency.  The interesting case is when $Q$ has a floor value, so we
consider that now. Figure \ref{sfeffic} shows the physical time,
$tV/R$, when $M_{\rm gas}=M_{\rm star}$, versus the star formation
efficiency, $S_0$, on the left and the $\sigma/V$ ratio at this time
versus $S_0$ on the right. The solid curves are for a floor value of
$Q=1$ and the dotted curves are for a floor value of $Q=0.7$ (beyond
which $\sigma$ increases proportional to $M_{\rm gas}$, as discussed
above). The different colors are for different turbulence-driving
efficiencies, $\epsilon$. Both $tV/R$ and $\sigma/V$ decrease with
higher $S_0$, and they diverge for low $S_0$. The divergence at low
$S_0$ is because the star formation rate per unit gas mass is so slow
with $Q$ at its floor value that the accretion keeps the gas mass
larger than the stellar mass, and our $M_{\rm gas}=M_{\rm star}$
condition for this figure cannot be satisfied. Also at low $S_0$, when
the time for $M_{\rm gas}=M_{\rm star}$ is large, the total gas mass
becomes large for the fixed accretion rate and $\sigma/V$ also has to
be large to keep $Q$ constant. There is a sharp turnaround in
$\sigma/V$ for $S_0>0.4$ (depending on $\epsilon$) with the $Q>1$
curves. In this region, the star formation efficiency is so high and
$M_{\rm gas}=M_{\rm stars}$ so early, that the turbulence is still from
gas accretion and $Q$ has not reached its floor value of unity yet.
Increasing $S_0$ beyond 0.4 increases $\sigma/V$ because the time gets
shorter in the accretion-dominated turbulent regime. These $Q>1$ curves
blend smoothly with the $Q>0.7$ curves at $S_0>0.4$ (depending on
$\epsilon$) because the $Q$ floor is not reached in either case for
large $S_0$ and then the $Q$ floor does not matter.

\section{Time-increasing Potential}
\label{time}

The basic equations are modified slightly if the galaxy potential
builds up with time along with the gas. If we suppose that the
potential is $V^2\left(t/t_0\right)$ instead of $V^2$ for constant $V$,
then equation \ref{edotprime} becomes
\begin{equation}
{{d{E^\prime}}\over{d{t^\prime}}}=t^\prime-M^\prime_{\rm
disk}\left({E^\prime}{M^\prime_{\rm
gas}}\right)^{0.5}-{S^\prime}{E^\prime/}{M^\prime_{\rm
gas}}\label{edotprime2}\end{equation} where
\begin{equation}
t_0={{\epsilon^{0.25}R}\over{A^{0.5}V}}\end{equation} is the same
factor that normalizes $t^\prime$ compared to $t$ ($t^\prime=t/t_0$).
All of the other equations are the same because they do not involve the
potential directly, except for $Q$, which becomes
\begin{equation}
Q^\prime  = \left(3t^\prime\right)^{0.5}\sigma^\prime/M^\prime_{\rm
gas}\end{equation} because the square root of the potential appears in
the epicyclic frequency, $\kappa$.

Two solutions of these and the other equations given above are shown in
Figure \ref{infall2f_variablepot}, which should be compared with Figure
\ref{infall2f}. One solution, with the solid curves, is for a growth of
the potential until $t^\prime=1$, at which point the galaxy mass is the
same as in Figure \ref{infall2f}, and then after this the galaxy mass
remains constant (i.e., the former equations are used again). The
second, shown by dashed curves in Figure \ref{infall2f_variablepot}, is
for a continuous growth up to arbitrarily large galaxy mass, which uses
these new equations for all $t^\prime$. The first case is a lot like
the previous solution except for $t^\prime<1$, where now the velocity
dispersion, $\sigma^\prime$, starts at zero instead of $\sim0.8$.  As a
result, the Jeans mass is very small throughout this solution and the
instability parameter $Q^\prime$ also starts small.  The other changes
are minor, especially for the bottom row of both figures, which assumes
$Q$ has a minimum value of 1 and plots unprimed quantities with
$A=0.08$ and $\epsilon=0.1$.

In the second case (dashed lines) where the potential continues to
grow, the normalized velocity dispersion stays large for a long time
(see top left panel of Fig. \ref{infall2f_variablepot}).  The rotation
speed is large too at these late times, higher than before by the
factor $\left(t^\prime\right)^{0.5}$.  The increase in $\sigma^\prime$
causes the star formation rate to increase more slowly than for a
constant potential, and the gas mass then gets higher before it equals
the stellar mass. The other changes are minor for this second case.

The changes to Figure \ref{infall2g} for a time-changing potential (not
shown) are qualitatively the same as the changes to Figure
\ref{infall2f} that are shown in Figure \ref{infall2f_variablepot}. The
velocity dispersion and $Q^\prime$ start low and the Jeans mass is low
throughout. The cases when $Q$ has a minimum value of 1 are hardly
changed at all.

Figure \ref{sfeffic}, which showed the time and relative velocity
dispersion when $M_{\rm gas}=M_{\rm star}$ in the $Q\geq1$ case, is
also hardly changed when the potential increases with time. This is
characteristic for $Q\geq1$.  However, the ratio of $\sigma$ to the
full rotation speed is now $\sigma/\left(V[t/t_0]^{1/2}\right)$ instead
of $\sigma/V$. This change can be significant, as seen in Figure
\ref{sfeffic_variablepot_sigmaonly}. The left-hand panel is for an
increasing potential up to $t=t_0$ and then a constant $V$. The right
hand panel is for a continuously increasing potential, with no limit.
The curves in the left panel are almost the same as in Figure
\ref{sfeffic} because the time when $S_0\lesssim0.3$ is larger than
$t_0$ and then the galaxy mass is the same as before. Recall that
$t_0=\epsilon^{0.25}R/\left(A^{0.5}V\right)$, so
$t/t_0=(A^{0.5}/\epsilon^{0.25})(tV/R)\sim0.50(tV/R)$ for $A=1/4\pi$
and $\epsilon=0.1$. This means that the time $tV/R$ plotted in the left
hand panel of Figure \ref{sfeffic} should be multiplied by 0.5 to
convert to $t/t_0$. The curves in the right panel have lower values of
$\sigma/V$ because the mass is larger at larger times, and so the
rotation speed is larger, when $S_0$  is small. The implication of this
change is that observed ratios of dispersion to rotation speed for
galaxies with $M_{\rm gas}=M_{\rm stars}$ permit smaller star formation
efficiencies if the galaxy mass continuously increases during gas
accretion.  The primary reason is that relative dispersions of several
tenths can occur later in galaxies with $M_{\rm gas}=M_{\rm stars}$ if
the galaxy mass is larger.

\section{Comparison with Observations}

High redshift galaxies are not yet observed at the phase that is most
relevant to this paper, namely, when turbulence is still driven by gas
accretion and star formation has not yet begun. Selection effects limit
the observation of galaxies to the starburst phase, at which point a
significant amount of turbulence should be driven by the dynamics that
triggers the star formation, most likely gravitational instabilities,
and also by the young stars themselves. Thus the fast turbulence
observed in young galaxies so far is probably not from accretion
energy, but from gravitational instabilities and young stars. However,
the turbulence that was present before the giant star-forming clumps
form, which caused the disk Jeans mass to be so large and gave these
clumps their enormous masses, could have been accretion energy. We
predict that observations of extremely young, gas-dominated disks will
show high turbulent speeds even before star formation begins.

\begin{figure}\epsscale{1}
\plotone{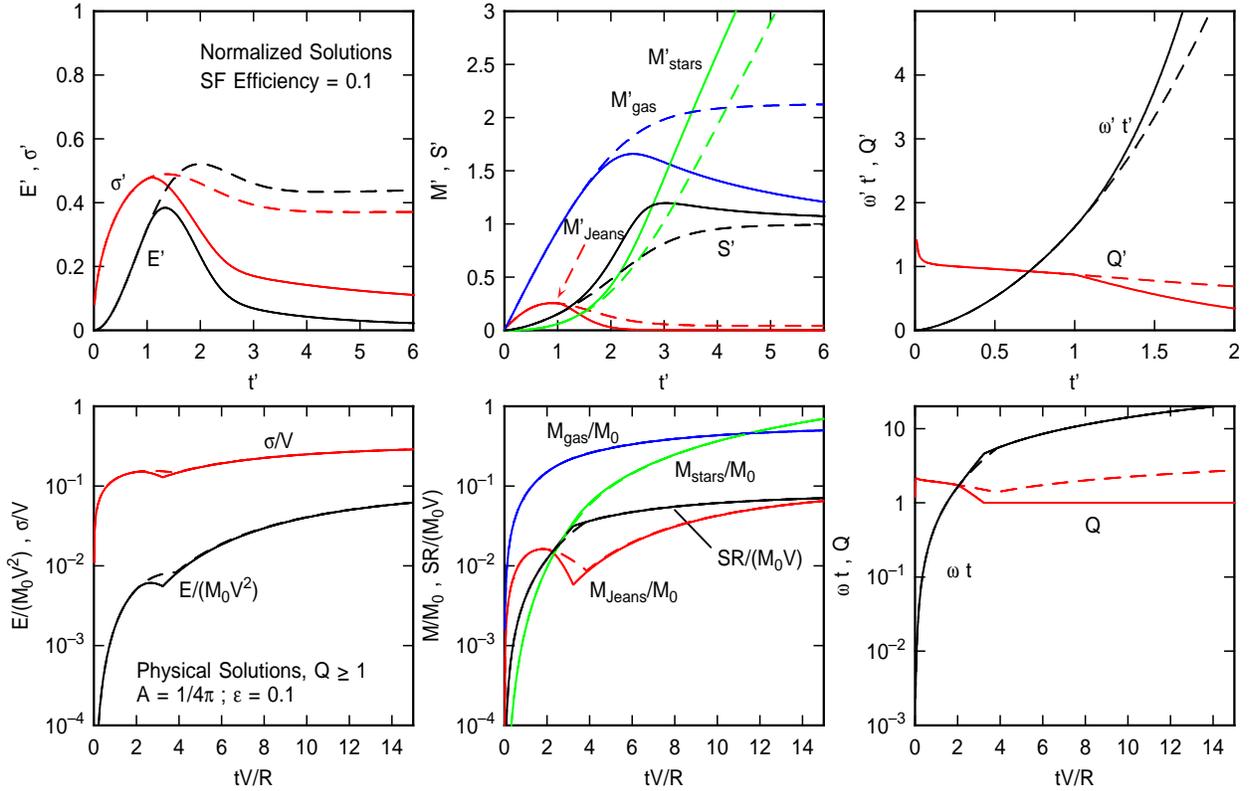} \caption{Same as Figure 1, but with higher star
formation efficiency, $S_0=0.1$. With higher efficiency, the stellar
mass increases sooner and faster, reaching equilibrium with the
accretion rate earlier. High $S_0$ is required to get $M_{\rm gas}\sim
M_{\rm stars}$ and large $\sigma/V$, as observed, when the velocity
dispersion is regulated by $Q\geq1$.}\label{infall2g}\end{figure}

\begin{figure}\epsscale{1}
\plotone{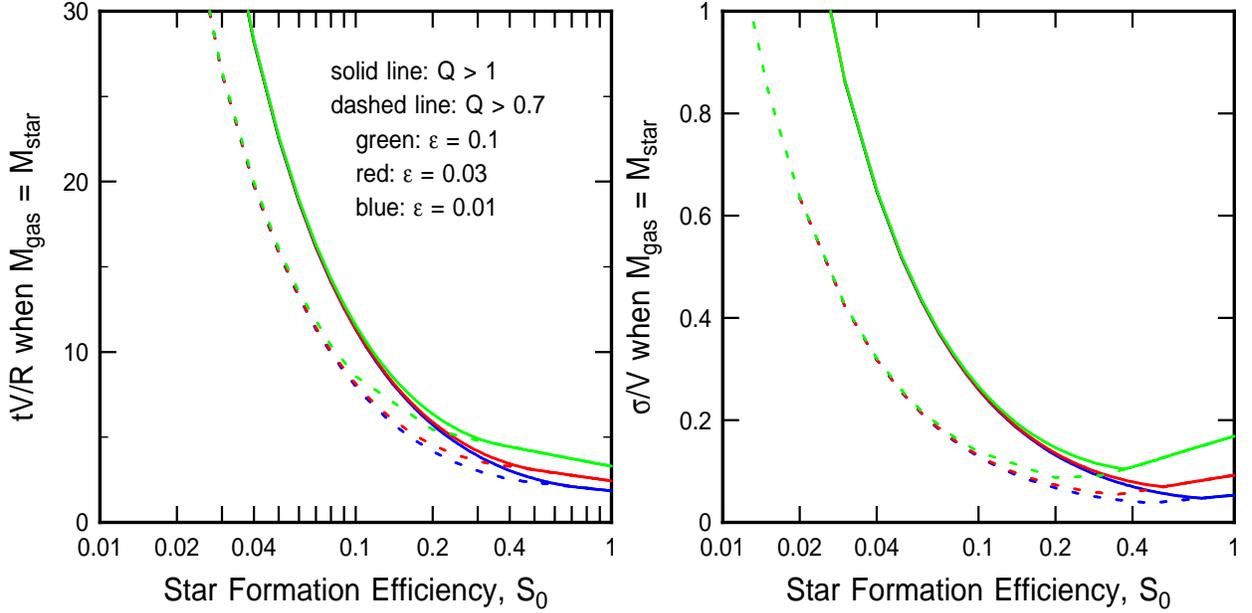} \caption{The relative time, $tV/R$, (left) and
relative velocity dispersion, $\sigma/V$, are shown versus the star
formation efficiency, $S_0$, at the time when $M_{\rm gas}=M_{\rm
stars}$. This epoch of mass equality is considered because it is
representative of disk galaxies at $z\sim2$. In all of these solutions,
the $Q$ parameter has a floor value (1 or 0.7), which is enforced by
making $\sigma$ increase in proportion to $M_{\rm gas}$. For low
efficiencies, the star formation rate cannot keep up with the accretion
rate, so $M_{\rm stars}$ is never equal to $M_{\rm gas}$. For high
efficiencies, stars build up quickly, decreasing the time when a
significant mass of stars appears, and decreasing the gas mass at that
time, from which the lower $\sigma/V$ follows.  The turn-around in
$\sigma/V$ at high $S_0$ is from the occurrence of $M_{\rm gas}=M_{\rm
stars}$ before the $Q$ floor is reached; elsewhere, $M_{\rm gas}=M_{\rm
stars}$ after the $Q$ floor is reached.}\label{sfeffic}\end{figure}

\begin{figure}\epsscale{1}
\plotone{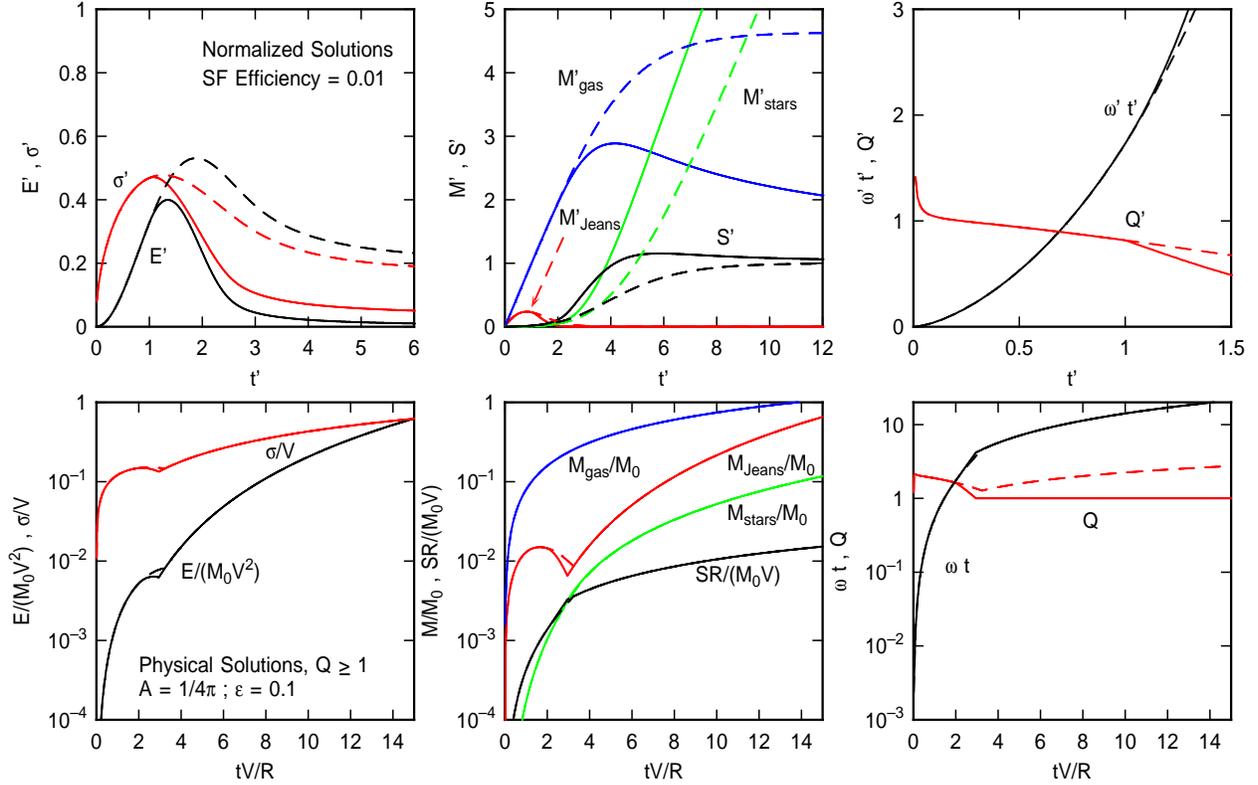} \caption{Models analogous to those in Figure 1 but now
with a potential that increases linearly with time. The solid curves
are for a case where the potential grows only until a dimensionless
time $t^\prime=1$, at which point the galaxy mass remains the same as
in Figure 1.  The dashed curves are for a case where the potential
continues to grow for all time. The primary effect of a growing
potential is to decrease the velocity dispersion, Jeans mass, and
stability parameter $Q$ at small times. The lower panels, which assume
$Q$ is regulated to remain larger than 1, are not significantly
affected by the change.} \label{infall2f_variablepot}\end{figure}

In the simple model presented here, accretion energy causes the
turbulent motions during the first few rotation times for a steady,
high accretion rate. In general terms, this phase lasts for $t_{\rm
acc}=2\epsilon^{0.25}R/(A^{0.5}V)$ considering that $t^\prime\sim2$ at
this time (from Figs. \ref{infall2f} and \ref{infall2g}). Writing
$A=\xi R/(VM_0)$, the duration becomes $t_{\rm acc} =
2\left(\epsilon^{0.5}RM_0/[\xi V]\right)^{0.5}=
2\left(\epsilon^{0.5}R^2V/[\xi G]\right)^{0.5}=2.2\left(\epsilon^{0.5}
GM_0^2/[\xi V^3]\right)^{0.5}$ for accretion rate $\xi$ in physical
units. Thus the accretion-driven turbulence phase lasts longer for
higher efficiency $\epsilon$ and lower accretion rate, and for more
massive or larger galaxies at a given rotation speed. For the typical
parameter values discussed elsewhere, $t_{\rm acc}\sim180 $ Myr.

The normalized time when the disk becomes gravitationally unstable,
i.e., $Q=1$, which means $Q^\prime=A^{0.5}/\epsilon^{0.25}=0.5$ for
$A=0.08$ and $\epsilon=0.1$, is found to be $t^\prime\sim1.7$ ($1.64$
in Fig. \ref{infall2f} with $S_0=0.01$ and $1.74$ in Fig.
\ref{infall2g} with $S_0=0.1$). These are for the cases with no floor
in $Q$. The normalized gas mass at this time is $M^\prime_{\rm
gas}\sim1.6$, (1.62 for $S_0=0.01$ and 1.53 for $S_0=0.1$) which means
the physical gas mass is $M_{\rm gas}=M^\prime_{\rm
gas}A^{0.5}\epsilon^{0.25}M_0=0.25M_0=2.3\times10^{10}\;M_\odot$ for
this normalization.  This is the gas mass at the beginning of the
starburst phase.  By the time the gas and stellar masses are equal,
which is appropriate for the current gas observations (e.g., Tacconi et
al. 2009), $t^\prime=3$ to 5 (5.5 for $S_0=0.01$ and 3.12 for
$S_0=5.5$) and $M^\prime_{\rm gas}=2.7$ to 1.6, respectively, making
$M_{\rm gas}\sim3.9$ to $2.3\times10^{10}\;M_\odot$.  With a $Q=1$
floor and $S_0=0.1$ (bottom panel of Fig. \ref{infall2g}), $tV/R=11.6$
when $M_{\rm gas}=M_{\rm stars}$, and then $M_{\rm
gas}=0.46M_0=4.1\times10^{10}\;M_\odot$.

These gas masses are appropriate for the start of the star formation
phase, between the time of first instabilities and the time when the
stellar mass has built up to be comparable to the gaseous mass. The
masses are large in our model because the young disk is stabilized by
turbulence that is driven by accretion energy. This energy source
explains how the gas mass can build up to such large values without
first turning into stars.  The youngest disks require stability like
this or else the gas-to-star ratio will always be low. We obtain the
observed high ratio and the observed gas masses for reasonable values
of the accretion rate and efficiency of turbulence driving. We also
obtain the high Jeans masses of the first star-forming events, which is
also an indication of fast turbulence.

Our conclusion that accretion dominates turbulence only during the very
earliest phase of galaxy growth may be derived most simply from the
dimensionless equations in Section \ref{acc-diss}. From equation
\ref{edotprime}, we see that energy gains exceed energy dissipation
when $1>M^\prime_{\rm disk}\left({E^\prime}{M^\prime_{\rm
gas}}\right)^{0.5}$. From equation \ref{mdiskdotprime}, $M^\prime_{\rm
disk}=t^\prime$. Also at early times, there are not many stars and
equation \ref{mdotprime} gives $M^{\prime}_{\rm gas}\sim t^\prime$.
Similarly, from the first term in equation \ref{edotprime},
$E^\prime\sim t^\prime$. From all of these we derive
$t^\prime\lesssim1$ when turbulence is driven by accretion. This
translates to a physical time
$t<t_0=\epsilon^{0.25}R/\left(A^{0.5}V\right)$, or, after substituting
$A=\xi R/(VM_0)$, we get $t^2<\epsilon^{0.5}RM_0/\left(\xi V\right)$,
as derived above.

Dekel, Sari \& Ceverino (2009) consider disk turbulence driven by
accretion, disk instabilities, disk clump stirring, and star formation.
For the accretion-dominant phase, they derive a timescale in their
equation 40, which is $\sim M_{\rm disk}\sigma^2/\left(\xi V^2\right)$.
This is essentially the same as we get because we both assume equality
between the turbulent energy and the accretion energy in this phase.
The simple derivation in the previous paragraph eliminates $\sigma$ in
favor of global galactic quantities because $\sigma$ is what we wish to
determine. Most of the discussion in Dekel et al. is about disk
self-regulation, which is more simply treated in the present paper by
assuming the instability parameter $Q$ has a minimum value of order
unity. Self-regulation occurs at a later phase, $t^\prime>1$, when
accretion is not strong enough to drive turbulence in comparison to
internal disk processes and star formation. Observations so far see
primarily this late phase.  Lehnert et al (2009) comment on
accretion-driven turbulence after observing spectra of redshift
$z\sim2$ galaxies. They note that the observed turbulence is too
dissipative to be driven by a reasonable accretion rate and conclude
that star formation dominates instead. Our model agrees with their
assessment, as the starburst phase occurs much later than the
accretion-driven turbulence phase, at $t^\prime\gtrsim3$ (Fig.
\ref{infall2f}).

Klessen \& Hennebelle (2009) suggest that turbulent accretion can power
turbulence in a wide variety of conditions, including the outer parts
of modern galaxies where the star formation rate is low. For high
redshift galaxies, they conclude that the observed turbulence inside
individual clumps can be driven by clump gas accretion if the total
rate is $10-50\;M_\odot$ yr$^{-1}$. They consider this reasonable as it
is comparable to the observed star formation rate. However, they also
caution that their model might not apply to the whole disk of a clumpy
high redshift galaxy, as clump coalescence and minor mergers might
dominate accretion and star formation in that phase.

\section{Conclusions}

Gas accretion to a young galaxy disk can drive turbulence for a time
$t_{\rm acc}\sim2\left(\epsilon^{0.5} GM^2/\xi V^3\right)^{0.5}$ where
$\epsilon$ is the fraction of the accretion energy going into disk
turbulence and $\xi$ is the accretion rate. After this time, disk
turbulence should be driven by gravitational instabilities and star
formation.  The first instabilities should produce only a few giant gas
clumps that dominate early star formation until the relative gas
fraction in the disk decreases.  The star formation rate equilibrates
to the accretion rate in 1 or 2 $t_{\rm acc}$, depending on the star
formation efficiency ($S_0$).

\begin{figure}\epsscale{1}
\plotone{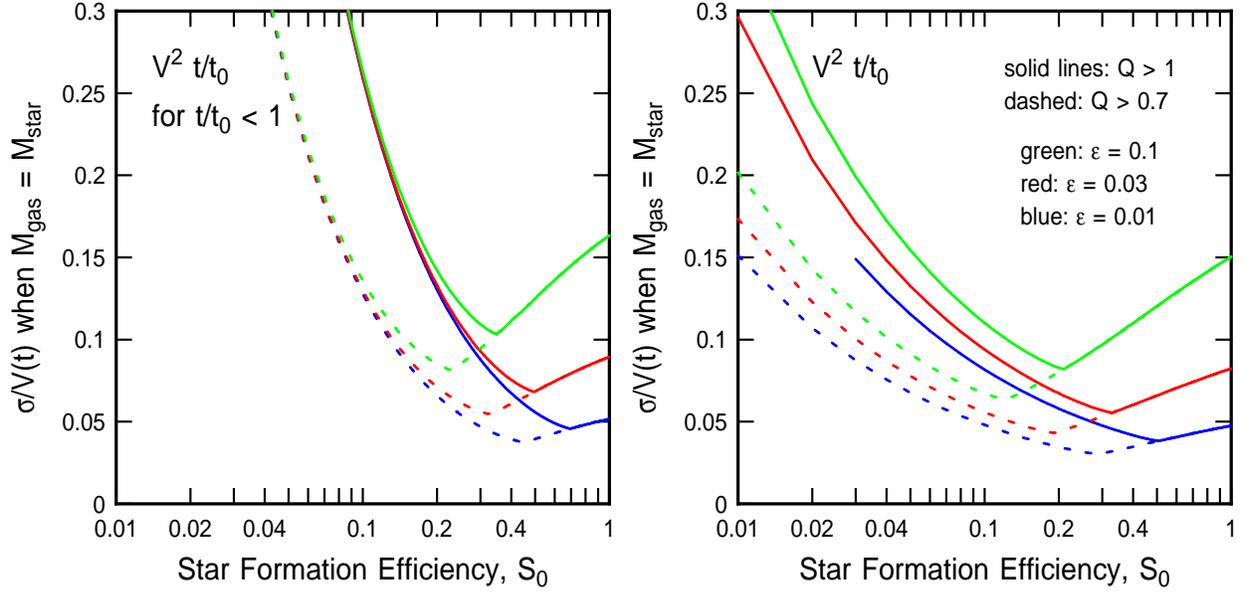} \caption{The ratio of the dispersion to the
instantaneous rotation speed is shown at the time when $M_{\rm
gas}=M_{\rm stars}$, as in Figure 5, but now with a time-increasing
potential. The solid and dashed lines are the same as before ($Q>1$ and
$Q>0.7$ self-regulation, respectively), and the three cases for the
efficiency of conversion of accretion energy into turbulent energy,
$\epsilon$, are also the same.  On the left, the potential increases
only until $t^\prime=1$ after which the mass remains the same as in
Figure 5. On the right, accretion continues for all time. The left-hand
panel is not changed much from before, but the right-hand panel
suggests that at low $S_0$, which corresponds to long times,
$\sigma/V(t)$ is lower when the potential continues to increases than
when the potential was constant (Fig. 5) or stops increasing at
$t^\prime=1$ (left panel). This is because the rotation speed is large
when the potential is large.}
\label{sfeffic_variablepot_sigmaonly}\end{figure}

Observations of a relative high turbulent speed compared to the
rotation speed, and of a disk gas mass comparable to the disk stellar
mass, require that the galaxy-averaged star formation efficiency has to
be large compared to the modern galaxy-averaged star formation
efficiency. This efficiency is defined here as the star formation rate
per unit dynamical time in the disk, and per unit total galaxy area.
Thus a high average value means either that the local efficiency is
high and the gas area equals the galaxy area, or the local efficiency
is more normal and the gas area is smaller than the galaxy area. Our
one-zone model cannot distinguish between these possibilities. The high
efficiency causes the stellar mass to build up to the gas mass before
the velocity dispersion in a $Q=1$ gas disk has dropped below a few
tenths of the rotation speed.  This criterion is easier to satisfy if
the total galaxy mass increases with the gas mass (Fig.
\ref{sfeffic_variablepot_sigmaonly}).

The high rate of star formation in young galaxy disks is the result of
a rapid instability compared to the galaxy age and a high gas mass
compared to the galaxy mass. The rapid instability follows from our
simple model (Fig. 3). The instability growth rate, $\omega$, compared
to the rotation rate, $V/R$, is $\omega R/V\sim\left(M_{\rm
gas}/M\right)\left(V/\sigma\right)$. This is large because the relative
gas mass,  $\left(M_{\rm gas}/M\right)$, is large for young disks, even
though the relative velocity dispersion, $\left(\sigma/V\right)$ is
somewhat large too.   As $\left(\sigma/V\right)$ drops because of
turbulent dissipation, and the accretion continues, $\omega V/R$
becomes even larger. The star formation rate declines only when the
accretion rate declines.

We conclude that gas accretion is a good source of energy for ISM
turbulence in the earliest phases of galaxy growth. It is quickly
replaced by other sources after a few rotation times, and is not likely
to be the source of turbulence that is observed at high redshift in
starbursting systems.   Accretion-driven turbulence in a young disk is
important because it provides the initial stability that allows the gas
to build up to a large mass before star formation begins. It also gives
the observed large masses for star-forming clumps. Observations of
gas-rich disks before the starburst phase should show the predicted
high turbulent speeds that come from accretion.

We are grateful to the referee for helpful suggestions.

\clearpage

\clearpage

\clearpage

\clearpage

\clearpage

\clearpage

\clearpage

\end{document}